\documentstyle[prb,aps,preprint]{revtex} % 23 pages preprint style
\begin{document}
\draft
\title{Finite-Temperature Elasticity Phase Transition in Decagonal
 	Quasicrystals}
\author{Hyeong-Chai Jeong and Paul J. Steinhardt}
\address{
  Department of Physics, University of Pennsylvania \\
  Philadelphia, Pennsylvania 19104
}
\date{Submitted to PRB}
\maketitle

\begin{abstract}
We present evidence for a novel finite-temperature phase transition in
the phason elasticity of quasicrystals. A tiling model for
energetically stabilized decagonal quasicrystals has been studied in
an extensive series of Monte Carlo simulation.
Hamiltonian (energetics) of the model is given
by nearest-neighbor Penrose-like matching rules
between three dimensional unit cells. A
new order parameter and diagnostics have been introduced. We show that
a transition from locked phason to unlocked phason dynamics occurs at
finite temperature. In the unlocked phase, phasons can be
thermodynamically excited even though the quasicrystal is
energetically stabilized at low temperatures.
\end{abstract}

%s1================================================================
\section{Introduction}

	Quasicrystals exhibit two distinct types of low-energy elastic
( hydrodynamic ) excitations --- phonons and phasons \cite{lev}.  In
this paper we consider the temperature-dependent behavior of the
phason elasticity in three-dimensional quasicrystals with decagonal
symmetry: a solid that can be described as a stack of periodically
spaced planes which each exhibit ten-fold symmetry \cite{ben}.  We
present evidence for a novel finite-temperature phase transition in
the phason elasticity, apparently analogous to the pinning transition
found in the one-dimensional Frenkel-Kontorova model (FK model) \cite{bak}.

	Quasicrystals are new types of solids which have a discrete point group
symmetry that is forbidden for crystals such as five-fold symmetry in
two-dimensions and icosahedral symmetry in three-dimensions \cite {lev1}.
These quasicrystals possess a long range translational order known as
quasiperiodicity.
The recent experiments on AlCoCu showed the existence of
thermodynamically stable decagonal quasicrystals \cite {kor}.
What makes the quasicrystals stable? Two possibilities that have been debated
are energetic stability and entropic stability \cite {ste}.
In energetically stabilized quasicrystal model, microscopic
interaction energy has its minimum when atoms are
arranged in a quasiperiodic structure. Such interactions ensure that
the low-temperature equilibrium state is quasicrystalline.
In entropically stabilized quasicrystal
model, the entropy that arises from thermodynamically excited
 atomic relocations (specifically, phasons)
makes the quasicrystal stable. In this model, quasicrystals are not
stable at low temperature.

	In this paper, we consider the phason dynamics of
energetically stabilized tiling model for the decagonal phase.
The interactions are prescribed so that quasicrystals remain stable as
the temperature T approaches zero. A tile is an idealization of
presenting a cluster of atoms in a real material. The energetics are
mimicked by nearest neighbor matching rules which are generalizations
of the Penrose edge-matching rules for two-dimensional tilings.
We assign a finite energy to each mismatch in a given configuration.
This guarantees that the state of lowest energy is a perfect
quasicrystal. This is to be contrasted with a random tiling model, used
to represent the limiting case of entropically stabilized
quasicrystals, in which the same energy is assigned to every
configuration and a quasicrystal symmetry is favored due to the high entropy.

	Phonons and phasons are low energy hydrodynamic modes
associated with quasiperiodic broken translational symmetry. At long
wavelengths, phonons correspond to uniform translations, and phasons
correspond to rearrangements of atoms from one perfect quasicrystal
lattice to another. In a tiling picture of quasicrystals, the phason
degrees of freedom correspond to the rearrangements of tiles. Finite
wavelength phasons produce rearrangements which violate the matching
rules and hence cost finite energy. If ${\bf w}(x)$ is the phason field,
then fixed phason strains produce a number of mismatches proportional
to $ |\Delta {\bf w}| $. Consequently, the elastic energy is
$ F \sim |\Delta {\bf w}|$, a nonanalytic form. We shall call a finite
temperature state in which the elastic energy has this form a
``locked phase'' \cite{hen}, since the phasons cannot be
thermodynamically excited (no phason Debye-Waller contribution).
An alternative type of quasicrystal is described by the continuum density
wave picture in which the elastic free energy grows $F \sim (\Delta
{\bf w})^{2}$. This density
wave picture corresponds to a distinct elastic phase which we shall refer
to as the ``unlocked phase''. In this phase, phasons have thermal
excitations analogous to phonons. Historically, the locked phase has
been assumed
to be characteristic of any energetically stabilized quasicrystal, whereas
entropically stabilized quasicrystals are, by definition, in the unlocked phase
with square gradient elasticity. Our point here is to show that this
view is not correct in general. Rather it is possible in
energetically stabilized quasicrystals to have a novel elastic phase
transition from a locked phase at low temperature to unlocked phase at
high temperature as had been speculated by Socolar {\it et. al.} \cite{soc}.
We note that this behavior is different from 2D where both energetic
and entropic model have unlocked elasticity at any finite temperature
\cite{{kal},{tan}}.
Hence, the observation of unlocked elastic behavior (phason Debye
Waller effect ) at finite temperature is not a proof of entropic
stability.

	The organization of the paper is as follows.  In sec. II,
we define our model for studying decagonal phase quasicrystals.
In sec III, we present the
characteristics of the system we use for Monte Carlo simulation. The
algorithms of Monte Carlo runs are also presented.  Sec. IV presents
diagnostics for studying the phase transition in phason elasticity.
We introduce
the notion of ``lane width'' and ``trail magnetization'' to check
if the system is in the locked phase.
The results of the Monte Carlo simulation are presented here.
Sec. V, the conclusion, summarizes our results and discusses the
potential connection to the experiment.

%2=================================================================
%s2.tex
\section {Defining Models}

	Our model for 3D quasicrystals with
decagonal symmetry has two types of unit cells: skinny and fat.
Each cell is the shape of a prism with rhombic
cross-section whose upper and lower faces are the shape of Penrose
rhombuses. In the zero temperature ground state, each layer viewed
along the ten-fold axis resembles a perfect Penrose tiling. The 2D
Penrose edge rules are replaced by ``side face'' rules, referring to
tile faces that join tiles in the same layer.
Side-faces of the unit cells have single or
double arrow marks as shown in Fig. 1 so that the tiling consistent
with matching rules $-$ joining side-face to side-face according to the
same arrow marks, is the perfect, quasiperiodic Penrose tiling.

We assign mismatch energy $\epsilon_{1}$ for a violation of single arrow
marks. Double arrow mismatches are not allowed (mismatch energy for
double arrow mismatch is infinity) in order to study purely phason
dynamics(see discussion below Eq. (3)). The 3D decagonal model is
introduced by considering a stack of these layers. We fix
the interlayer interaction energy to have
its minimum when configurations of two layers are identical (fat tile
directly on fat tile, skinny directly on top of skinny).
We assign stacking direction face mismatch energy $\epsilon_{2}$ when
the vertices of an upper[lower] face of a unit cell do not coincide
with those of a lower[upper] face of an upper[lower] unit cell.
The energy of a tiling is defined as
a sum of intralayer interaction energy (arrow mismatch energy) and
interlayer interaction energy (stacking direction face mismatch
energy):
\begin{eqnarray}
E & = & E_{\mbox{intralayer}} + E_{\mbox{interlayer}}   \nonumber \\
  & = & \sum_{\mbox{side faces}} \varepsilon_{\mbox{intra}}
               + \sum_{\mbox{upper, lower faces}} \varepsilon_{\mbox{inter}}
\end{eqnarray}
where
\[ \varepsilon_{\mbox{intra}} = \left\{ \begin{array}{ll}
                           0    & \mbox {for an arrow matched side face} \\
            \epsilon_{1} & \mbox {for a single arrow mismatched side face} \\
            \infty & \mbox {for a double arrow mismatched side face,} \\
                          \end{array}
                  \right.
\]
and
\[ \varepsilon_{\mbox{inter}} = \left\{ \begin{array}{ll}
               0 & \ \mbox{ for an upper[lower] face which coincides with} \\
                 & \ \mbox{ lower[upper] face of an upper[lower] unit} \\
  \epsilon_{2} &  \ \mbox{ for an upper[lower] face which does not coincide} \\
                 & \ \mbox{ with lower[upper] face of an upper[lower] unit.} \\
                    \end{array}
                  \right.
\]
Henceforth we will refer to intralayer mismatches as ``arrow
mismatches'' and interlayer mismatches as ``face mismatches''.

	In the finite temperature Monte Carlo, we introduce
`` flips '' of single hexagonal prisms within a layer.  There are two
kinds of hexagonal prisms which we shall call D-type hexagonal prisms
and Q-type hexagonal prisms, respectively. A D[Q]-type hexagonal prism
consists of two fat cells and one skinny cell [one fat cell and two
skinny cells] and its top view is a D[Q]-type hexagon \cite{bru}.
Each hexagonal prism can be uniquely identified by the vertex at the
center of the upper hexagon. Fig. 2 shows hexagonal prism before and
after ``flips''. Solid circles indicate the center vertices of the upper
hexagons.

A hexagonal prism flip from a ground state makes two arrow mismatches and 6
face mismatches. Hence, the net energy cost is
$ 2 \epsilon_{1} + 6 \epsilon_{2} $.

	For most of our MC runs, we impose a ``stacking
constraint'':
For a given hexagonal prism, flipping is allowed only when there is a
hexagonal prism in the adjacent upper [or lower] layer whose
lower[upper] hexagon has a boundary that coincides (ignoring center
vertex) with the boundary of upper[lower]
hexagon of the given hexagonal prism.
This constraint is necessary for ``interlayer flips'' \cite{{hen},{shi}}.

	Let us define phason variables in this model. The position of
any vertex of the tiling with fixed origin (by choosing a vertex as
the origin) is expressed as
\begin{equation}
{\bf x} = \sum_{\alpha=0}^{5} n_{\alpha} {\bf e}^{\parallel}_{\alpha},
\end{equation}
where
$ {\bf e}^{\parallel}_{\alpha} = ( \cos \frac{2\pi \alpha}{5},
\sin \frac{2\pi \alpha}{5}, 0 )
$ for $\alpha = 0, 1, \cdots, 4 $ and (0, 0, 1) for $\alpha = 5$,
$n_{\alpha}$ is the number of steps in direction ${\bf e}^{\parallel}_{\alpha}$
on a continuous path to the vertex at ${\bf x}$ from the origin along the
edges, counting negatively when going in the $-{\bf e}^{\parallel}_{\alpha}$.
We define a complementary basis ${\bf e}^{\perp}_{\alpha}$, such that the
vectors
${\bf e}_{\alpha} = {\bf e}^{\perp}_{\alpha} \oplus {\bf
e}^{\parallel}_{\alpha} $ are linearly independent in 6D
hyperspace. Then, for each vertex ${\bf x}$ whose position is given by Eq.
(2), we define a perp-space position vector
$ {\bf x}^{\perp} $ by
\begin{equation}
	{\bf x}^{\perp} =  \sum_{\alpha=0}^{5} n_{\alpha} {\bf e}^{\perp}_{\alpha}
\end{equation}
where $ {\bf e}^{\perp}_{\alpha} = ( \cos \frac{4\pi \alpha}{5}, \sin
\frac{4\pi \alpha}{5}, 1 )
$ for $\alpha = 0, 1, \cdots, 4 $ and (0, 0, 0) for $\alpha = 5$.
These vectors span a 3D space called `perp-space' which can be further
decomposed as the
product of a 2D-`phason space' (which is incommensurately
oriented with respect to the hyperspace lattice) and an 1D-`discrete space'
(which is commensurately oriented). In our model in which double arrow
mismatches are disallowed, the discrete space component of ${\bf x}^{\perp}$ is
restricted to four consecutive integers. Phason variables ${\bf w}$
are to be defined as a smoothed function ${\bf w}({\bf x})$, constructed
as an average of the phason space projection ${\bf x}^{\perp}_{ph}$ of the
perp-space
position vector ${\bf x}^{\perp}$,
in some neighborhood of radius $a_{0} \gg 1$.
%s3================================================================
\section {Monte Carlo Simulation}
	We use a thermal Monte Carlo methods
based on Metropolis importance sampling scheme \cite{mon}.
The basic Monte Carlo move is: \
 1) Randomly select a hexagonal prism. \
 2) if it satisfies the stacking direction constraint, flip it
according to a probability; \ \
$ p = \exp (-\beta \Delta E) \ $ if \ $\Delta E > 0$,  \ \ $ p = 1$
\ otherwise, \ where $\Delta E$ is the energy cost for performing the flip.
	We used an approximant to a 2D perfect Penrose tiling for a
layer which has the smallest phason strain at a given system size with
periodic boundary conditions. To get a periodic tiling, we use a
periodic uniform dual grid which is the intersections of 4D lattice
planes of the 5D hypercubic lattice with a 2D hypersurface spanned by
two vectors,
\begin{equation}
\vec{\omega}_1 = ( \ P, \ 0, -P, -Q, \ Q),
\ \ \vec{\omega}_2 = ( \ 0, \ P, \ Q, -Q, -P),
\end{equation}
where P and Q are integers.
A sequence of approximants are obtained by taking
$  P = F_{k}, \ Q = F_{k-1}, $
where $F_{k}$ is $k$th Fibonacci number $(F_{0} = 0, \ F_{1} = 1)$.
Then the basis vectors of the periodic tiling are given by
\begin{equation}
 {\bf L}^{(1)} = \tau^{k} \ (\frac{2\tau-1}{2},
             \ -\frac{1}{\tau}\sin\frac{\pi}{5}),
\ \ {\bf L}^{(2)} = \tau^{k} \ ( 0, \ 2\sin\frac{\pi}{5}),
\end{equation}
and the uniform phason strain ${\bf E}$ to get the $k$th periodic
approximant is
\begin{equation}
 {\bf E} = (-1)^{k+1} \tau^{-2k} \left( \begin{array}{cc}
 				1 & \ \ \  0 \\
             2 \sin\frac{2\pi}{5} & \ \ \ \tau \\
                              \end{array}
                              \right) .
\end{equation}
When we calculate perp-space position vectors, to get rid of the
effects due to the uniform phason strain ${\bf E}$,
we replace ${\bf e}^{\perp}_{\alpha}$ in Eq. (3) by
${\bf e}^{\perp}_{\alpha} - {\bf E} {\bf e}^{\parallel}_{\alpha}$ .
The number of tiles in a layer is given by $N_{k} =
F_{2k+1} + 2 F_{2k}$ . Since our main interest is the properties of 3D
objects, the number of layers $L_{z}$ of the systems is
proportional to the size of a layer $(L_{z} = F_{k}$ for the $k$th
approximant). Systems of \ 228($N_{k} \times L_{z} = 76\times 3$),
\ 995($199\times5$), \ 4168($521\times8$) and 17732($1364\times13$)
tiles are used.
We also study effect of varying $L_{z}$ for fixed
$L^{1}$ and $L^{2}$.
For most of the models presented here, we have assigned
interaction energy strength $\epsilon_{1}=1$ and
$\epsilon_{2}=1/3$ in Eq. (1), corresponding to equal interlayer and
intralayer energy cost for flipping an isolated hexagon beginning
from the ground state.

	We start out with an ordered configuration, with the minimum
matching rule violations necessary to construct the  periodic approximants.
Data are taken following a heating sequence. We studied the time
evolution of the data to check whether the system has been
equilibrated. To ensure statistical independence, we measured the
temporal correlations and most case measurements are separated more
than correlation time \cite{sha}.  Some cooling runs have been
performed from above $T_{c}$ to below $T_{c}$ to check the presence of
hysteresis effect. The data so obtained agree in high accuracy with
those from heating studies.
As a test of the algorithm, we assigned
$\epsilon_{2} = 0$ without the stacking direction constraint and get
the known 2D Penrose model \cite{tan} results.
%s4================================================================
\section  {Diagnostics for phason elasticity transition and Results of
	Monte Carlo simulation }
	In the unlocked phase, where the continuum density wave picture is
applied, the general form of the phason elastic free energy (up to
quadratic in $\partial {\bf w}$) is
\begin{equation}
	F = \frac{1}{2} \int \ d^{d}x K_{ijkl} \partial_{i} w_{j}
\partial_{k} w_{l},
\end{equation}
 where $w_{j}$ is the $j$th component of phason variable and
$K_{ijkl}$ is an elastic constant tensor. For our model, due to
the decagonal symmetry, the free energy of Eq. (7) reduces to the form
\begin{eqnarray}
 F &= \frac{1}{2} \int \ d^{3}x &  \ \ \ K_{1}
  [ (\partial_{x} w_{1})^{2} +  (\partial_{y} w_{1})^{2}
  + (\partial_{x} w_{2})^{2} +  (\partial_{y} w_{2})^{2} ]  \nonumber \\
   & &  + K_{2}  [ (\partial_{z} w_{1})^{2} +  (\partial_{z} w_{2})^{2} ]
 + K_{3}  [ \partial_{x} w_{1} \partial_{y} w_{2}
          - \partial_{y} w_{1} \partial_{x} w_{2} ]
\end{eqnarray}
where $K_{1}, K_{2}$ and $K_{3}$ are elastic constants in unlocked
phase. The integration of the last term in a layer can be dropped to a
line integral along the boundary of the layer that vanishes in the
absence of dislocations.

	On the other hand, in locked phase the phason elastic free
energy is given by
\begin{equation}
 F = \frac{1}{2} \int \ d^{3}x  \ \ \ \tilde{K_{1}}
  ( |\nabla_{\perp} w_{1}| +  |\nabla_{\perp} w_{2}|)
  + \tilde{K_{2}} ( |\partial_{z} w_{1}| +  |\partial_{z} w_{2}| )
\end{equation}
where $\tilde{K_{1}}$ and $\tilde{K_{2}}$ are elastic constants in
locked phase and $\nabla_{\perp} = {\bf e}_{x} \partial_{x} + {\bf e}_{y}
\partial_{y}$ is two dimensional derivative in the plane perpendicular
to stacking direction (z-axis).

	To see whether a system is in a locked phase or in an unlocked phase,
we could, in principle, compare the free energies of the systems with
different uniform phason strain ${\bf E}$ .
To do this we would need to know the energy and
the entropy of the system with low uniform background phason strain
${\bf E}$, as functions of temperature. We attempted to estimate the entropy,
using the
`` energy method '', ``the variance method '' and ``the histogram
method'' \cite{{oxb},{swe}}, but we could not estimate entropy
accurately enough to distinguish the phason elasticity.
Hence we resorted several other measures including
phason fluctuations, lane widths and trail
magnetization(defined below).

%s4.a---------------------------------------------------------------
\subsection  {Phason Fluctuations}

The phason elastic free energy in the unlocked phase (8) can be
diagonalized by Fourier transforming of phason variable ${\bf w}({\bf x})$,
\begin{equation}
 {\bf w}({\bf p}) = V^{-1/2} \int d^{3} x \ e^{ -i{\bf p} \cdot {\bf x}} {\bf
w}({\bf x}) ,
\end{equation}
where V is the volume of the system.
Then free energy (8) becomes
\begin{equation}
 F = \frac{1}{2} \sum_{|{\bf p}|<\frac{2\pi}{a_{0}}}
   \  K_{1} (p_{x}^{2} + p_{y}^{2}) ( w_{1}({\bf p})^{2} + w_{2}({\bf p})^{2})
\  + \  K_{2} \ p_{z}^{2} \ ( w_{1}({\bf p})^{2} + w_{2}({\bf p})^{2}).
\end{equation}
(We ignore the third term in Eq (8) which vanishes in
any configuration in our simulation. )
Since the free energy (11) is
harmonic in $ w_{i}({\bf p})$ for $i = 1, 2 $, it is straightforward to
calculate $ <|{\bf w}({\bf p})|^{2}> $,
\begin{equation}
  <|{\bf w} ({\bf p})|^{2}>
 = \sum_{i=1,2} <|w_{i} ({\bf p})|^{2}>
 = \frac{2}{ K_{1} (p_{x}^{2} + p_{y}^{2}) + K_{2} p_{z}^{2}},
\end{equation}
where the angular brackets denote ensemble average.
Elastic constants $K_{1}$ and $K_{2}$ are obtained by measuring
$<|{\bf w}({\bf p})|^{2}>$ for ${\bf p} = p_{x} {\bf e}_{x} + p_{y}
{\bf e}_{y}$ and for ${\bf p} =  p_{z} {\bf e}_{z}$. Let us define

\begin{eqnarray}
 K_{1}({\bf p}) & \equiv & \frac {2}{ \ <|{\bf w}(p_{x}, p_{y}, 0)|^{2}>
             \ (p_{x}^{2}+p_{y}^{2})} , \nonumber \\
 K_{2}({\bf p}) & \equiv & \frac {2}{ \ <|{\bf w}( \ 0, \ 0, p_{z})|^{2}> \
 	    p_{z}^{2} \ .}
\end{eqnarray}

If the system is in the unlocked phase, $K_{1}({\bf p})$ and $K_{2}({\bf p})$
should be constants, independent of $ |{\bf p}|$ for $ \frac{2\pi}{L} < |{\bf
p}| <
\Lambda $ , where $L$ is the system size and $\Lambda$ is short wavelength
cutoff which depends on the coarse graining scheme ($\Lambda$ is order
of $\frac{2\pi}{a}$ for a smooth function ${\bf w}(x)$, where $a$ is size
of a unit cell).
A proper coarse graining scheme is necessary to get elastic constants
by measuring $<|{\bf w}({\bf p})|^{2}>$ in our simulation. This is
because the perp-space position ${\bf x}^{\perp}$ (from which a smooth function
${\bf w}(x)$
is constructed) fluctuates strongly for near-neighbor vertices
( $ \Delta {\bf x}^{\perp} / \Delta {\bf x} \sim 1 $ \ for \ $ \Delta {\bf x}
\sim 1 $).

We construct the phason field ${\bf w}({\bf x})$ from the perp-space position
${\bf x}^{\perp}$ by
\begin{equation}
{\bf w}({\bf x}) = \int d^{3} x \ K({\bf x} - {\bf x}') \ S^{\perp} ({\bf x}'),
\end{equation}
where   $K({\bf x})$ is a smearing kernel
satisfying $\int d^{3} x \ K({\bf x}) = 1$ and
$S^{\perp}({\bf x})$ is the representative surface $-$ piecewise linear
interpolating function of ${\bf x}^{\perp}_{ph}$.
Precisely, we use
\[
S^{\perp}({\bf x})  = \left \{ \begin{array}{ll}
	{\bf x}^{\perp}_{ph}({\bf x})
	& \mbox { for ${\bf x}$ at a vertex of the upper face of a cell,} \\
	{\bf x}^{\perp}_{ph}({\bf x}_{a}) d_{b} + {\bf x}^{\perp}_{ph}({\bf x}_{b})
d_{a}
	& \mbox { for ${\bf x}$ in an edge ($d_{a}$ from one end ${\bf x}_{a}$,} \\
	& \mbox { $d_{b}$ from the other end ${\bf x}_{b}$) of the upper
	 	face,} \\
	\frac{1}{2} (S^{\perp}({\bf x}_{1}) + S^{\perp}({\bf x}_{2}) )
	& \mbox { for ${\bf x}$ inside the upper face, where ${\bf x}_{1}$} \\
	& \mbox { and ${\bf x}_{2}$ are the projections of ${\bf x}$ to the } \\
	& \mbox { edges (meet at a vertex) of the upper face,} \\
	S^{\perp}({\bf x}_{up})
	& \mbox { for ${\bf x}$ inside the unit cell, where ${\bf x}_{up}$ is} \\
	& \mbox { the projection of ${\bf x}$ to the upper face} \\
                  \end{array}
                  \right.
\]
and
\begin{equation}
K({\bf x}) = \frac{1}{4\pi} \frac{1}{r_{xy}} \Theta(2-r_{xy}) \delta(z),
\end{equation}
where $r_{xy} = (x^{2}+y^{2})^{1/2}$ and $ \Theta(x) $ is the step
function. When this coarse graining scheme is applied to the 2D Penrose
model, the known elastic constants \cite{tan} are recovered for
$|{\bf p}| < \Lambda \simeq 1.5$.
Note that a simple-minded coarse graining scheme to replace Eq. (15),
${\bf w}({\bf x}) =
\frac{V}{N} \sum_{i=1}^{i=N} {\bf x}^{\perp}_{ph}({\bf x}_{i}) \ \delta({\bf x}
- {\bf x}_{i})
\  ({\bf w}({\bf p}) = \frac{V^{1/2}}{N} \sum_{i} {\bf x}^{\perp}_{ph}({\bf
x}_{i}) \ e^{ -i {\bf p} \cdot
{\bf x}}) $
would not have worked as well. In the 2D Penrose model,
the known elastic constants \cite{tan} are recovered for only
$|{\bf p}| < \Lambda \simeq 0.2$ which is order of long wavelength cutoff
($\frac{2\pi}{L}$) of the biggest system in our simulation.

	To determine if the system is in the unlocked phase, we
measure whether $K_{1}({\bf p})$ and $K_{2}({\bf p})$ as defined in Eq. (13)
are $|{\bf p}|$ - independent. In Fig. 3, we show a plot of $K_{1}({\bf p})$
and $K_{2}({\bf p})$ vs $|{\bf p}|$ at $T=1.5$ and 2.5 for a sequence of
lattices of increasing size. These plots show that $K_{1}({\bf p})$ and
$K_{2}({\bf p})$ are constants and independent of system size, indicating
that the system is unlocked at $T > 1.5$ . We estimate the elastic
constants to be
$K_{1} = 1.72 \pm 0.02 , \ K_{2} = 0.56 \pm 0.02$ at $T = 1.5$ and
$K_{1} = 1.22 \pm 0.02 , \ K_{2} = 0.14 \pm 0.01$ at $T = 2.5$.

	In contrast, Fig. 4 illustrates the same calculation for
$T=1.0$ . Here, $K_{1}({\bf p})$ and $K_{2}({\bf p})$ fluctuate wildly with
$|{\bf p}|$ and the mean value appears to diverge with increasing system
size. Hence $T=1.0$ is clearly below a phason transition (out of the
unlocked phase). The wild behavior of  $K_{1}({\bf p})$ and $K_{2}({\bf p})$ is
consistent with the notion that the phason elastic energy is locked.
Hence, phase fluctuation measurements can be used to establish a
transition out of the unlocked phase described by an elastic energy of
the form in Eq. (8). However we cannot prove from phason fluctuation
measurements that the low temperature phase has $F \sim |\nabla {\bf w}|$ (as
expected for a Penrose tiling phase). This is because the phase
fluctuations appear to become pinned (as indicated by the divergence
of $K_{1}$ and  $K_{2}$ ). Since ${\bf w}$ is not thermodynamically
excited, the dependence of $F$ on $|\nabla {\bf w}|$ cannot be measured.

	Further evidences of the transition from the unlocked phase
is  provided by the measurement of the average phason field within a
layer:
\begin{equation}
 \overline{\bf w}(z) = \frac{1}{S} \int_{S} {\bf w}(x,y,z) \ dx dy,
\end{equation}
(where $S$ is the area of a layer) and then calculating how this
average fluctuates from layer to layer.
The average phason field within a layer $\overline{\bf w}(z)$ is related to the
Fourier components of phason field at $p_{x} = p_{y} = 0$ by,
\begin{equation}
 \overline{\bf w}(z) = V^{-\frac{1}{2}} \sum_{p_{z}} {\bf w}(p_{x}=p_{y}=0
,p_{z}) e^{ip_{z}z},
\end{equation}
since
\begin{eqnarray}
 {\bf w}(p_{x}=p_{y}=0,p_{z})
   &=& V^{-\frac{1}{2}} \int {\bf w}(x,y,z) e^{-ip_{z}z} \ dv \nonumber \\
   &=& (\frac{S}{L_{z}})^{\frac{1}{2}} \int \overline{\bf w}(z) e^{-ip_{z}z} \
dz ,
\end{eqnarray}
where $L_{z}$ is the number of layers. In an unlocked phase
$|{\bf w}(p_{x}=p_{y}=0, p_{z})|^{2} = 2/ (K_{2} \ p_{z}^{2})$ (Eq. (12)).
Hence, the mean square fluctuation of $\overline{\bf w}(z)$ in the unlocked
phase:
\begin{eqnarray}
 <(\Delta \overline{\bf w})^{2}>
  & \equiv &  \frac{1}{L_{z}} <\sum_{z} |\overline{\bf w}(z) -
	     \frac{1}{L_{z}} \sum_{z} \overline{\bf w}(z)|^{2}>  \nonumber \\
   &=& \frac{1}{V}\frac{L_{z}}{2\pi} \int_{c/L_{z}}^{c/a}
       \frac{2}{K_{2} \ p_{z}^{2}} \ dp_{z} \nonumber \\
   &=& \frac{2}{2 \pi S c K_{2}} ( L_{z} - a ),
\end{eqnarray}
where c is order of $2\pi$ and $c/a$ is upper wave number cutoff.
Fig. 5 shows $ <(\Delta \overline{\bf w})^{2}> $ vs $L_{z}$ at $T = 1$, $T=1.5$
and $T=2.5$. The initial configuration within each layer is a 6th
approximant. $ <(\Delta \overline{\bf w})^{2}> $does not show a linear
dependence
on $L_{z}$ at $T=1.0$ while it
is linear in $L_{z}$ at $T=1.5$ and $T=2.5$ (consistent with
the earlier conclusion of an unlocked phase at $T=1.5$ and $T= 2.5$).
{}From the slope in Fig. 5.b and Eq. (19), we find $cK_{2} = 1.10 \pm
0.06$ at $T=1.5$ and  $cK_{2} = 0.27 \pm .03$ at $T=2.5$.
Comparing these values to the elastic
constants from the measurements of $<|{\bf w}({\bf p})|^{2}>$, implies cutoff
constant $c \sim 2$ in Eq. (19) while we have $ a \simeq 1$ from
Fig. 5.

%s4.b---------------------------------------------------------------
\subsection {Lane widths}
	As a mean of analyzing the low-T phase, we have measured the
spacing between two adjacent trails in a layer.

	A ``trail'' in our model is a contiguous strip of tiles which share
a common side face direction (side face direction ${\bf q}_{\alpha}$ of a side
face
parallel to the plane spanned by ${\bf e}^{\parallel}_{\alpha}$ and ${\bf
e}_{5}^{\parallel}$ is
$ {\bf q}_{\alpha} = {\bf e}^{\parallel}_{\alpha} \times {\bf
e}_{5}^{\parallel}, \ \alpha = 0, \ldots, 4 $).
We shall call a trail which has a side face direction ${\bf q}_{\alpha}$, an
$\alpha$-direction trail. An $\alpha$-direction trail runs along the
${\bf q}_{\alpha}$-direction on average. As shown in Fig. 6, each layer
consists of
sets of parallel trails (Fig. 6 shows a layer in a 5th approximant
system viewed from the ten-fold axis).
The regions between the trails will be referred to as ``lanes''.
In a perfect Penrose tiling, two different widths of lanes exist:
thick lane and thin lane. These lanes repeat quasiperiodically
(Fibonacci sequence) in the direction normal to the trail direction.
Hence, in a locked phase, where this quasiperiodicity is believed not
to be destroyed, the distribution of the lane widths is bimodal.
The distribution under the one mode (corresponding thick
lanes) is $\tau$ times bigger than that under the other mode
where $\tau$ is Golden
mean.  In the unlocked phase, the distribution of the lane widths may have
merged into  one peak (as the distribution of the distances between nearest
balls
in a unlocked phase of Frenkel-Kontorova model\cite{bak}). With this in
mind, we have measured the number of lanes $N^{lane} P(W)dW$ whose
average width is in between $W$ and $W+dW$ where $N^{lane}$ is the
total number of lanes in the system (average width of a lane is defined
as the average of the spacing between two trails contiguous with
the lane (see Fig. 6)). At low-$T$ (at $T$ = 1.0), as shown in
Fig. 7.a, $P(W)$ has two peaks near the values corresponding the
lane widths ($W_{1}^{lock}$ and $W_{2}^{lock}$) of a Penrose tiling.
As the system size diverges,
lane widths converge the values of Penrose tiling lane widths.
In contrast, $P(W)$ at high-$T$ (at $T$=1.5, Fig. 7.c) shows a
distribution with one peak at $W^{unlock}=L/N^{lane}_{\alpha}$ as
$L\rightarrow \infty$, where $L$ is the system size and
$N^{lane}_{\alpha}$ is the number of $\alpha$-direction lanes in a
plane ($N^{lane}_{\alpha} = f_{k}$ for a $k$th approximant).
$P(W)$ near the transition temperature (at $T=1.3$) is shown in Fig. 7.b.
{}From this figure, it is hard to tell whether the curve $P(W)$ at
$T=1.3$ will be bimodal (as in a locked phase) or be monomodal (as in
a unlocked phase) as $L \rightarrow \infty$.
We have checked reversality by doing some cooling runs from above $T_{c}$.
The distribution of the lane widths merged into one peak at high-T,
come back to original bimodal mode with peaks at Penrose tiling lane
widths.
This measurement of the distributions of the lane widths, suggests
that the system is in locked phase at low temperature.

	To estimate the transition temperature $T_{c}$ we have
measured the number of lanes whose widths are around the lane widths
of a Penrose tiling and lanes with widths near $W^{unlock}$.
Let us define $P^{lock}$ and $P^{unlock}$ as the relative number of
these lanes:
\begin{eqnarray}
P^{lock}   & \equiv & \int_{R_{1}} P(W) dW + \int_{R_{2}} P(W) dW,
			\nonumber \\
P^{unlock} & \equiv & \int_{R_{3}} P(W) dW, \nonumber \\
P          & \equiv & P^{lock} - P^{unlock},
\end{eqnarray}
where
 $ R_{1} \equiv (W_{1}^{lock} - \delta_{1}, W_{1}^{lock} + \delta_{1}),
 \ R_{2} \equiv (W_{2}^{lock} - \delta_{1}, W_{2}^{lock} + \delta_{1}), $
and
 $R_{3} \equiv (W^{unlock} - \delta_{2}, W^{unlock} + \delta_{2}).$
Here, $\delta_{1}$ and $\delta_{2}$ are small numbers ($\ll 1$) with
$\delta_{2} = 2 \delta_{1}$.
Then $P$ should be 1 at $T=0$ and should be negative at $T = \infty$
(the precise value at $T = \infty$ depends on the choice of
$\delta_{1}$ and $\delta_{2}$ but greater than $-1$ always).

Fig. 8. shows $P$ vs temperature for a sequence of lattices of
increasing size. We have chosen $\delta_{1} = 0.05$ and $\delta_{2} =
0.1\ . \ \ P$ converges to 1 at low $T$ and has negative values at high $T$.
We roughly estimate the transition temperature
$T_{c} \sim 1.3$ from Fig. 8 as the value around which
the graphs of different sizes cross each other.

%s4.c---------------------------------------------------------------
\subsection { Trail Magnetization}
	As a new order parameter to analyze the low-$T$ phase and
the transition from the low-$T$ phase,
we have devised a novel measure that we have termed ``
trail magnetization''. Trail magnetization traces the ordering of
``hexagonal prisms'' along worms and trails that run through each layer.

	A hexagonal prism, as shown in Fig. 2, consists of three tiles.
The type (D or Q) of a hexagonal prism
is identified with the type of the center vertex
on the upper hexagon of the hexagonal prism \cite{bru}. By convention, the
orientation of a hexagonal prism is ``+''[``$-$''] if the center vertex
has an edge leading away from it along direction
${\bf e}^{\parallel}_{\alpha}$ [$-{\bf e}^{\parallel}_{\alpha}$].
 A worm in a layer is an unbroken sequence
of hexagonal prisms and the length of a worm is defined to be the
number of consecutive, connected hexagonal prisms.
Flipping one hexagonal prism in a perfect worm creates a mismatch along either
side face adjoining the worm ( flipping the hexagonal prism
again annihilates the
mismatches). If we restrict flipping to only one worm, the 2D
Penrose model is analogous to the one dimensional Ising model,
assigning `spin up' for one orientation hexagon and `spin down' for the
flipped hexagon in a worm.
If the 2D Penrose tiling were an aggregation of uncoupled worms (
{\it i.e} uncoupled one dimensional Ising models), we would expect an
order-disorder transition at $T=0$ which coincides with the result
that the transition to the unlocked phase in 2D Penrose tiling is at
$T=0$.

 Along a trail, many worms( chains of hexagonal prisms) may be found. Along
any trail in a perfect Penrose tiling, all worms longer than one
hexagonal prism have the same orientation. Different trails may have
opposite hexagonal prism orientation. Also, some worms of length one
point in the opposite direction to the other hexagons in a trail.
All together, about 97\% of hexagonal prisms in a trail have
same orientation in a perfect Penrose tiling while the hexagonal prism
in a maximally random tiling are oriented randomly.
Hence, we define the trail magnetization in $\alpha$ orientation
\begin{equation}
  m_{\alpha}^{tr} = \frac{1}{N_{\alpha}^{hexa}} \sum_{i \in \{tr_{\alpha}\}}
		| \sum_{j \in tr_{\alpha}^{i}} S_{j}|,
\end{equation}
where $N_{\alpha}^{hexa}$ is number of hexagons in $\alpha$-direction trails,
and $\{tr_{\alpha} \}$ means all trails ($f_{k} \times L_{z}$ trails
for a $k$th approximant) in $\alpha$-direction
and $tr_{\alpha}^{i}$ means the $ith$ trail in $\alpha$-direction ($i
= 1, \ldots, f_{k} \times L_{z}$).
Here, spin variable $S_{j}$ is
assigned to the  $j$th hexagonal prism and takes $\pm 1$ depend on
the orientation of the hexagonal prism.
	Without loss of generality, we will discuss results for $\alpha
= 0$ (${\bf e}_{0}^{\parallel}$-direction trails).

	Figure 9 shows the trail magnetization and
 ``trail susceptibility''
\begin{equation}
 \chi_{m} = N_{hexa} \ \ \frac{1}{T} (<(m^{tr})^{2}> - <m^{tr}>^{2}),
\end{equation}
and illustrates how $m^{tr}$ can serve as a useful diagnostic measure.
At $T=0$, the trail magnetization converges to a fixed value as the
system diverges, a value consistent with the expectation value for a
locked, Penrose tiling phase.
At $T \geq 1.5$, the trail magnetization approaches zero as $L
\rightarrow \infty$, consistent with an unlocked phase.
The magnitude of susceptibility maximum seems to
diverge near the transition temperature.
To obtain the transition temperature and critical exponents, we
attempt to fit the trail magnetization and susceptibility according to
\begin{eqnarray}
 m^{tr}(T,L)  & = & L^{-\beta/\nu} f[(T-T_{c}) L^{1/\nu}], \\
 \chi_{m}(T,L) & = & L^{\gamma/\nu} g[(T-T_{c}) L^{1/\nu}],
\end{eqnarray}
where $L$ is the system size, $T_{c}$ is transition temperature of
the infinite system and critical exponents $\beta, \ \gamma, \ \nu$ are
defined from the temperature dependence of order parameter $m^{tr}$,
susceptibility $\chi_{m},$ and the correlation length $\xi$ near the
transition temperature:
\begin{eqnarray*}
 m^{tr}   & \sim & (T-T_{c})^{\beta}, \ \ \ \mbox {for} \ \ (T < T_c) \\
 \chi_{m} & \sim & |T-T_{c}|^{-\gamma}, \\
 \xi      & \sim & |T-T_{c}|^{-\nu}. \\
\end{eqnarray*}
In Fig. 10.a, we plot $m^{tr}(T, L) \ L^{\beta/\nu} $ vs
$(T-T_{c}) L^{1/\nu} $ and in Fig. 10.b,
$\chi_{m}(T, L) \ L^{-\gamma/\nu}$ vs $(T-T_{c}) L^{1/\nu} $,
for values
\begin{eqnarray}
   T_{c}   =  1.24 \ \   & \ \ \beta = 0.2   \nonumber \\
   \gamma  =  1.6  \ \ \ & \ \ \nu   = 1.6.  \nonumber
\end{eqnarray}
The $m^{tr}$ and $\chi_{m}$ curves for various system size
superimpose clearly.
We have checked the possibility of a sequence of transitions rather
than a single transition by considering
\begin{equation}
  {m'}_{\alpha}^{tr} = \frac{1}{N_{\alpha}^{hexa}} \sum_{i \in \{tr_{\alpha}\}}
		|\sum_{j_{l} \in tr_{\alpha}^{i}} S_{j_{l}}|,
\end{equation}
where  $tr_{\alpha}^{i}$ traces over the $ith$ trails in all layers along
direction $\alpha$.
(To see the ordering in stacking direction also,
we took a sum of spin valuables in $\ ith$ trails of all $L_{z}$ layers
before taking the absolute values
instead of summing the spin values in each individual trail as in $m^{tr}$.)
We have been able to make
${m'}^{tr}$ and $\chi_{m'}$ curves for various system size superimpose.
The transition temperature from the measurement of ${m'}^{tr}$ is
consistent with the result from $m^{tr}$.

	We also measured energy per tile $<\varepsilon> = <E/N>$ and specific
heat $ C_{v} = (\frac{1}{T})^{2} (<\varepsilon^{2}> -
<\varepsilon>^{2}) N$, where $E$ is the system energy given by Eq. (1)
and $N$ is number of tiles. Specific heat has its maximum near the
transition temperature $T_{c}$ and the magnitude of its maximum seems
to be independent of the system size (Figure 11) implying that the specific
heat exponent $\alpha = 0$ where $\alpha$ is defined by,
\begin{eqnarray}
C_{v}     \sim  |T-T_{c}|^{-\alpha}.  \nonumber
\end{eqnarray}

The results of trail magnetization and specific heat measurements
seem very consistent with a single continuous transition
from a locked phase ($m^{tr} \approx 1$) to an unlocked phase.

	To check the robustness of the result,
we have repeated the analysis for other related models.
First we changed the interaction strength ratio
$\epsilon_{1}/ \epsilon_{2}$. Recall that the net energy cost of a
hexagonal prism flip is $ 2 \epsilon_{1} + 6 \epsilon_{2} $.
We have tested $\epsilon_{1}/\epsilon_{2} = 1$ and
$\epsilon_{1}/\epsilon_{2} = 9$ and found that the systems show a
locked phason to unlocked phason phase transition at finite
temperature.

	We have also considered models where we have relaxed the
interlayer ``stacking constraint'' (that a hexagonal prism can be
flipped only if a hexagonal prism lies just above or below with
sharing a same hexagon boundary in between; see sec. II).
The low temperature phase of this model appears to be the same as that of the
model with the stacking constraint. At high temperature ($T
> T_{c} $), free energy shows a quadratic dependence on spatial
variation in phason variable in a layer ($ |\partial_{x} {\bf w}|^{2} +
|\partial_{y} {\bf w}|^{2} $). For the stacking direction,
free energy shows a quadratic behavior in  ($ \partial_{z} {\bf w} $)
up to some temperature $T'_{c} \ (T'_{c} > T_{c}
) $ which depends on the system size. Above the $T'_{c}$, layers become
decoupled so that the averaged phason field $\overline{\bf w}(z)$ of $z$th
layer
(Eq. (16)) is not correlated with $\overline{\bf w}(z \pm 1)$ (correlation
length
of $\overline{\bf w}(z)$ is zero for $T>T'_{c}$). However $T'_{c}$ increases as
the
system size is getting bigger and we speculate that $T'_{c}$ may diverge
in the thermodynamic limits.

%s5================================================================
\section  {Conclusions}

Our paper presents systems that show a finite temperature locked
phason to unlocked phason phase transition in quasicrystals. All
models for 3D decagonal quasicrystals we have studied show a single
continuous phase transition at finite temperature from low-$T$
locked phase to high-$T$ unlocked phase.

	Phason fluctuations of the system show strong evidences that
the system is in unlocked phase at $T>T_{c}$ in which free energy is
described as square gradient of phason variables. At $T<T_{c}$, from
the measurements of ``lane width'' and ``trail magnetization'' we
conclude that the system is in locked phase. The finite
specific heat peak and the critical exponents we obtained from the scaling
behaviors of the $m^{tr}$ show that the transition is continuous.

Recently Hiraga {\it et. al.}  have noted that the periodically
spaced layers in AlPbMn
decagonal quasicrystals strongly correlated atomic order.
Our simulations show that there is a single transition in which both
intralayer and interlayer phason fluctuations transform from locked to
unlocked.  Hence, the observation of correlated order (locking) between
layers would imply that
that AlPbMn quasicrystals are in the
locked phase and not in the random tiling in spite of some apparent
disorder within the layer.  (The disorder is likely to be due to decapod
defects or unusual local isomorphism class.)

	To predict the transition temperature $T_{c}$ in real quasicrystals,
we need to know how big the mismatch energies  ($\epsilon_{1} \ and
\ \epsilon_{2}$) are. Note that $T_{c}$ is of order the geometric mean
 mismatch energy $\epsilon = (\epsilon_{1} \epsilon_{2})^{\frac{1}{2}}$
for our model. $T_{c}$ will approach zero if either $\epsilon_{1}$
or $\epsilon_{2}$ approaches zero.

	The unlocking transition could be preempted by the melting transition.
If the transition temperature $T_{c}$ is
higher than the melting temperature $T_{m}$, quasicrystals remain in
the locked phase for temperatures ranging all the way up to melting point.
For these systems, one would expect more quenched phasons than for a
system which has unlocked phase between the melt phase and the locked
phase since unlocking implies rapid relaxation of phason fluctuations.
Hence, the relation between $T_c$ and $T_m$ may partially
account for the reason
why some systems form near-perfect quasicrystals and others do not.
For the system which has an unlocked phase ($T_{c} < T_{m}$),
the transition from the unlocked phase to the locked phase
could be observed in experiments.
An observational effect could be Debye-Waller suppression of the
diffraction peak intensities. In the unlocked phase, phasons can be
thermodynamically excited. As we can see in Fig. 3, the phason
elastic constants increase with decreasing temperature in the unlocked
phase. Consequently, the Debye-Waller suppression decreases as $T$
decreases toward $T_{c}$.  Then, after the transition to the locked
phase, thermal phason
fluctuations are frozen and the the Debye-Waller suppression disappears.

\acknowledgments
We would like to thank Tomonari Dotera for useful discussions.
This research was supported by the DOE at Penn (DOE-EY-76-C-02-3071).

%sbi===============================================================

\clearpage

%sf================================================================
%sf.tex
{\large Figure Caption} \\
{\bf Fig. 1 }
(a) Skinny Cell:
Upper and lower faces are in the shape of a Penrose skinny rhombus.
Side faces have arrow marks according to the arrow patterns
of the Penrose skinny rhombus. \\
(b) Fat Cell:
Upper and lower faces are in the shape of a Penrose fat rhombus.
Side faces have arrow marks according to the arrow patterns
of the Penrose fat rhombus. \\
{\bf Fig. 2} Hexagonal prisms: \\
(a) $D$-type hexagonal prism before ($+D$) and after ($-D$) flip. \
(b) $Q$-type hexagonal prism before ($+Q$) and after ($-Q$) flip. \
Solid circles indicate the center vertices of the upper hexagons. The
orientation ($\pm$ sign) is given by the convention explained in the
text (section IV.c). \\
{\bf Fig. 3}  The elastic constants $K_{1}({\bf p})$ (a) and
$K_{2}({\bf p})$ (b) defined by Eq. (13) vs the magnitude of the wave
vector $|{\bf p}|$ with unlocked phases (at $T=1.5$ and $T=2.5$).
$K_{1}({\bf p})$ and $K_{2}({\bf p})$ are constant over $|{\bf p}|$
and independent of the system size. In this and the following figures,
the numbers in the legend represent the numbers of tiles $N$ in the systems. \\
{\bf Fig. 4}  The elastic constants $K_{1}({\bf p})$ (a) and
$K_{2}({\bf p})$ (b) vs the magnitude of the wave
vector $|{\bf p}|$ with locked phase (at $T=1.0$).
$K_{1}({\bf p})$ and $K_{2}({\bf p})$ increase in
magnitude with increasing system size.
Points of data for each system size are connected by lines shown in
the legend.
The average of $K_{i}$ over $|{\bf p}|$ for
each system size is indicated by an arrow.\\
{\bf Fig. 5} Mean square fluctuation, $(\Delta \overline{\bf w})^{2}$,
of $\overline{\bf w}(z)$ is plotted vs
stacking direction size $L_{z}$ with locked phase (a)
and unlocked phases (b).\\
{\bf Fig. 6} Trails and lanes in a layer (above) and average widths of
lanes (below). The sequences of the shaded tiles are trails. The region
between neighboring trails is a lane. Average lane width is
defined as the average of the spacing between two trails contiguous with that
lane. \\
{\bf Fig. 7} Lane width distributions with locked phase (a) and
unlocked phase (c). \\ Lane width
distributions near the transition temperature are shown in (b). \\
{\bf Fig. 8} System size dependence of $P$ defined in Eq. (20) vs
temperature. \\
 {\bf Fig. 9} System size dependence of the trail magnetization (a)
and susceptibility (b) plotted against temperature. \\
{\bf Fig. 10} Finite-size scaling plots of the data for trail
magnetization, shown in Fig. 9.a, and for susceptibility, shown in
Fig. 9.b. \ Here, \ $ m L^{\beta/\nu} $ (a) \ and  \ $ \chi_{m}
L^{-\gamma/\nu} $ (b) \ are plotted versus $(T - T_{c}) L^{1/\nu}$
with the following choice of exponents: $\beta = 0.2, \ \gamma = 1.6,$
and $\nu = 1.6 $ and the transition temperature $T_{c} = 1.24$.\\
{\bf Fig. 11} System size dependence of specific heat plotted against
temperature.\\

\end{document}